\documentclass[10pt,conference]{IEEEtran} 
\usepackage[utf8]{inputenc}
\usepackage{cite}
\usepackage{graphicx}
\usepackage{hyperref}
\usepackage{multirow}
\usepackage{tabulary}
\usepackage{xcolor}

\newcommand{\keyword}[1]{\textit{#1}}
\newcommand{\code}[1]{\textsl{{#1}}}
\newcommand{\qq}[0]{\textquotesingle\textquotesingle}
\newcommand{\str}[1]{\qq{}{#1}\qq{}}

\newcommand{\safeds}[0]{Safe-DS}
\newcommand{\sklearn}[0]{scikit-learn}
\title{Safe-DS: A Domain Specific Language to \\ Make Data Science Safe}
\author{\IEEEauthorblockN{Lars Reimann}
\IEEEauthorblockA{Institute of Computer Science III\\
University of Bonn, Germany\\
Email: reimann@cs.uni-bonn.de}
\and
\IEEEauthorblockN{Günter Kniesel-Wünsche}
\IEEEauthorblockA{Institute of Computer Science III\\
University of Bonn, Germany\\
Email: gk@cs.uni-bonn.de}
}

\begin{document}

\IEEEoverridecommandlockouts
\IEEEpubid{
  \begin{minipage}{\textwidth}
  \noindent\rule[0.5ex]{0.5\textwidth}{0.75pt}\\
  Accepted at the 45th Int. Conf on Software Engineering, NIER Track\\
  ICSE-NIER’23, May 14–20, 2023, Melbourne, Australia\\
  \copyright2023 IEEE 
  \end{minipage}
}

\maketitle

\IEEEpubidadjcol

\begin{abstract}

Due to the long runtime of Data Science (DS) pipelines, even small programming mistakes can be very costly, if they are not detected statically. However, even basic static type checking of DS pipelines is difficult because most are written in Python. Static typing is available in Python only via external linters. These require static type annotations for parameters or results of functions, which many DS libraries do not provide.

In this paper, we show how the wealth of %elaborate 
Python DS libraries can be used in a \emph{statically safe} way via Safe-DS, a domain specific language (DSL) for DS. 
Safe-DS catches conventional type errors plus errors related to \emph{range restrictions}, \emph{data manipulation}, and \emph{call order of functions}, %Thus
going well beyond the abilities of current Python linters.
Python libraries are integrated into Safe-DS via a \emph{stub language} for specifying the interface of its declarations,
%in the Safe-DS API, 
and an \emph{API-Editor} that is able to extract type information from the \emph{code and documentation} of Python libraries, and automatically generate suitable stubs.
%, along with their implementing wrapper code. 

Moreover, Safe-DS complements textual DS pipelines with a graphical representation that eases safe development by preventing 
%typographical and 
syntax errors. 
%It makes it easy for programming novices to get started and eases communication of developers and non-programmers. 
The \emph{seamless synchronization} of textual and graphic view lets developers always choose the one best suited for their skills and current task. 

We think that Safe-DS can make DS development easier, faster, and more reliable, significantly reducing development costs. 

\end{abstract}


% ====================================================================
\section{Introduction}
\label{sec:introduction}

Various researchers have noted that the APIs of widely-used libraries are often hard to learn and use correctly \cite{Robillard:IEEESOftware2009,Robillard:ESE2011,ZibranEtAlWCRE2011,WangGodfrey:MSR2013}. 
%
%As shown by \cite{reimannImprovingLearnabilityMachine2022}, the APIs of widely-used data science (DS) libraries are no exception and  novices\footnote{That is, ``persons who aim to develop a DS application and have the required mathematical and software engineering background, but no experience using a particular DS API''. \cite{reimannImprovingLearnabilityMachine2022}} 
%are particularly affected. 
%
%Wrapping a library into a smaller and better designed API, reduces the cognitive burden for novices and helps \emph{statically} avoid errors such as mistyping string constants and passing integer values outside a documented legal range \cite{icse23, reimannImprovingLearnabilityMachine2022}. 
% 
% Detection of misuse is particularly crucial in DS: Programs that crash after running several hours or days for a reason that could have been detected at compile-time are a prohibitive waste of time, resources, and, ultimately, money. 
%
Prevention of erroneous use is particularly crucial in Data Science (DS): Programs that crash at run-time for a reason that could have been detected at compile-time are generally a waste of time, resources, and, ultimately, money -- and downright prohibitive after running hours or days. Therefore, static safety would be a major benefit for every Data Scientist, novice or master.
Unfortunately, most DS pipelines are written in Python, and use Python DS libraries \cite{kaggle2021KaggleMachine}. For Python, static type checking is only available via external linters like mypy\footnote{\url{https://github.com/python/mypy}}, which are, however, inapplicable if the code is not annotated with \keyword{type hints} \cite{TypingSupportType}.
Unfortunately, many libraries do not provide type hints, and no Python linter we know of is able to handle advanced type concepts 
like schema types for tabular datasets, 
which would help a lot in typical DS applications. 

% This way, errors such as mistyping string constants and passing integer values outside a documented legal range can be caught statically.} 
% OLD: Unfortunately, most DS pipelines are written using libraries implemented in Python \cite{kaggle2021KaggleMachine}, a language that is notoriously hard to type-check statically since it does not mandate the use of static type annotations \cite{TypingSupportType}. 
%for parameters and results of functions%, 
% Thus, many DS libraries do not provide them, which makes even basic static type checking within DS pipelines difficult.
% Therefore, Reimann et al. \cite{reimannImprovingLearnabilityMachine2022}, automatically analyze an API's documentation and, where possible, generate static type annotations.
% This way, errors such as mistyping string constants and passing integer values outside a documented legal range can be caught statically.
% Even if 
%static type 
% annotations are available, 
%Python simply ignores them, so 
%users must install external tools like mypy\footnote{\url{https://github.com/python/mypy}} to %benefit from 
%use them.

% it cannot catch more sophisticated classes of errors, for which no linters exist.  
%such as errors caused by (1) changes of the structure of datasets during data preparation, or by (2) using ML operations in the wrong order, e.g. using a model for prediction before it was trained. 

A radically different approach is taken by systems that hide the DS APIs behind a visual pipeline specification language (e.g. RapidMiner Studio \cite{mierswaYALERapidPrototyping2006}), a wizard-like interface (e.g. RapidMiner Go \cite{rapidminerGo}), or a configuration language (e.g. Ludwig \cite{ludwig}).
However, wizards and configuration languages can only provide guidance \cite{reimannAchievingGuidanceApplied2020a} and safety at the expense of limiting what can be expressed. Visual programming can offer full expressiveness, but when a pipeline gets complex, a graph representation quickly gets cluttered, and textual languages excel. Depending on the skills of the user and the use case, different representations are optimal.
However, no known system provides %both, 
visual \emph{and} textual ways to specify %entire 
DS pipelines. 
%, which is no surprise, because synchronizing these two views is non-trivial, given their different abstraction levels.
%A well-known example is synchronization of UML \cite{uml-spec} diagrams with code in common programming languages, e.g. Java. It is easy to map a high abstraction level (UML) to a low one (code) but hard to automatically determine which code snippets correspond to which abstraction.

%\subsection{Research Questions}
\label{sec:research-questions}

%This leads us to 
In this context, we explore the following research questions:
\begin{itemize}
     \item RQ 1: 
        How can we make the most widely used libraries for DS usable  even by programming novices?
        % people with limited programming skills?
    \item RQ 2: 
        Is it possible to design a language that is expressive enough to implement any DS pipeline but simple enough to be learned and used correctly, even by novices? 
        % \streichbar{Which features does it need and which must it avoid?}
    \item RQ 3: 
        Which classes of errors are not caught by conventional type systems (especially those of Python linters) but occur frequently in DS development?
    \item RQ 4: 
        Which %(common or advanced 
        type concepts could help and should be part of a dedicated DS language?
    \item RQ 5:
        How can a DS language ensure statically safe use of DS libraries written in dynamic languages?%, such as Python?
    \item RQ 6: 
        How can we make % our system extensible and 
        integration of existing Python DS libraries easy, even for third parties?
    % \item RQ 7: 
    %     How can we enforce type safety without burdening pipeline developers with specifying types?% in their programs? 
    \item RQ 7: 
        How can we leverage the built-in error prevention of graphical languages without sacrificing the flexibility of textual code? 
        %
        % \streichbar{Can we guarantee seamless synchronisation of both views to support users with varying expertise?}
\end{itemize}

% ====================================================================
%\subsection{Approach Overview}
%\label{sec:approach-overview}

%In this paper, we suggest a novel way to support the full stack of DS-relevant APIs while spotting most of the typical errors in their use already at compile-time: 
The solution suggested in this paper is \safeds{}, a statically typed domain specific language (DSL) \cite{enwiki:dsl} for specifying DS pipelines. It provides: 
\begin{enumerate}
    \item 
        A simple mechanism to integrate existing Python DS libraries in a type-safe way (Sec. \ref{sec:static-checking}). 
    \item
        A well-designed standard library that includes the best available DS libraries (briefly outlined in Sec. \ref{sec:prototype}). 
    \item 
        Support for third-party developers to extend the standard library by integrating their own features (Sec. \ref{sec:checking-external-code}).
 %   \item 
 %       a unique point at which all static checking can be expressed once and for all, benefiting all possible views of a DS pipeline (Sec. \ref{sec:static-checking}).
   \item 
        A seamless bidirectional transition between graphical and textual representation, which are just different views of the same abstraction, the DSL (Sec. \ref{sec:different-views}). 
\end{enumerate}

% ====================================================================
\section{State of the Art}
\label{sec:state-of-the-art}

According to the 2021 Kaggle DS survey \cite{kaggle2021KaggleMachine}, over 84\% of almost 26.000 respondents use Python regularly, making it the most commonly used DS programming language. 
According to the survey, the most-often used libraries for DS are 
Matplotlib \cite{Hunter:2007}, % (67.7\%)
Seaborn \cite{Waskom2021}, % (48.5\%)
and Plotly \cite{plotly} % (22.2\%)
for data visualization, 
Pandas \cite{The_pandas_development_team_pandas-dev_pandas_Pandas} for data processing, and 
\sklearn{} \cite{scikit-learn, sklearn_api}, %(53.9\%), 
TensorFlow \cite{tensorflow2015-whitepaper}, % (36.1\%), 
Keras \cite{chollet2015keras}, %(30.1\%), 
PyTorch \cite{NEURIPS2019_9015}, % (23.4\%)
and XGBoost  \cite{Chen:2016:XST:2939672.2939785} % (23.0\%).
for ML\footnote{Each group is sorted by descending usage percentages\cite{kaggle2021KaggleMachine}}.
All are implemented in Python, emphasizing its role in DS development. 

Nevertheless, there are noteworthy suggestions for DS pipeline development beyond Python:
Ludwig \cite{ludwig} allows users to build a DS pipeline with a YAML \emph{configuration file}. Users must specify input and output features, and can optionally alter some parts of the pipeline, such as data preprocessing.

External domain specific languages are another solution. The list of DSLs used for DS in 2016 \cite{portugalPreliminarySurveyDomainSpecific2016} includes \emph{textual DSLs} like Pig Latin \cite{olstonPigLatinNotsoforeign2008} for data processing and \emph{visual DSLs} like \cite{breukerModelDrivenEngineeringBig2014} for the representation of probabilistic models. Since then, other tools appeared, e.g. BiDaML \cite{khalajzadehBiDaMLSuiteVisual2019, khalajzadehBiDaMLPracticeCollaborative2021}, which offers five diagrammatic notations for different stages of the development of a DS pipeline. Another popular tool is RapidMiner Studio (formerly YALE) \cite{mierswaYALERapidPrototyping2006}, which offers a visual DSL to quickly, and % --- due to static checks --- 
safely build a DS pipeline.

Finally, some tools use \emph{wizards} to guide users step by step through the development of a DS pipeline, for example RapidMiner Go \cite{rapidminerGo}.
%These tools are generally less flexible than tools from the other categories but because of this easier to learn and use correctly. 
To overcome the limited flexibility of a wizard, users can export the final result to RapidMiner Studio.

\cite{keryFutureNotebookProgramming2020} extends Jupyter Notebook \cite{kluyverJupyterNotebooksPublishing} by additional GUI elements like a spreadsheet view to inspect data. When users manipulate the data in these elements, corresponding code is added to the notebook. However, changes to the generated code can in general not be reflected in the GUI, making the transition between graphical and textual view one-way only.

These ideas can be combined with \emph{AutoML} \cite{heAutoMLSurveyStateoftheart2021} to automatically derive parts of 
% or even an entire 
a DS pipeline from the initial data.
%Due to space constraints, we restricted our list to common examples for each category. 

An overview of free tools for data mining can be found in \cite{jovicOverviewFreeSoftware2014} and a more thorough survey of the DS landscape in \cite{nguyenMachineLearningDeep2019}. Even taking this larger context into account, no existing tool provides a comprehensive solution to 
% all our research questions.
the majority of our research questions, let alone all of them.

% \todo{
%     \begin{itemize}
%         \item Various DSLs to better optimize ML programs (e.g. parallelization)
%         \item Various commercial tools (Rapidminer, Amazon, Azure, Google, ...)
%         \item No code / low code solutions (configuration, graphical)
%         \item Dataiku
%         \item AWS
%         \item Azure
%         \item Google
%         \item Specialized tools for data transformation like Tableau
%         \item https://www.kdnuggets.com/2019/05/poll-top-data-science-machine-learning-platforms.html
%         \item Pre-trained models like Huggingface
%     \end{itemize}
% }

% ====================================================================
% Was hier auskommentiert stand ausgelagert nach "backup-roles.tex"
% ====================================================================

\section{A Language for Data Science Pipelines}
\label{sec:dsl}

A \emph{DS pipeline} is code that performs data loading, data preparation, model training, and hyperparameter optimization to produce a model that is ready for use. 
% When rerun on the same data, the pipeline will reproduce the same model. 
It conserves the steps necessary to reproduce the same model from the same data, without the data analysis and model quality assessment steps that developers performed to decide how to build the pipeline. 

The main aim of our work is to support pipeline developers of varying skill levels, including absolute programming and DS novices. For this, assembling DS pipelines should be easy to learn and use, but hard to misuse. Any solution must provide a small and consistent API (to ease learning and correct use), which must be easily extensible (to take advantage of the evolution of DS libraries). Extension must be possible and easy, also for third-party library developers.

% \begin{figure}[ht]
%     \centering
%     \includegraphics[width=.49\textwidth]{img/Safe-DS-users.png}
%     \caption{Safe-DS supports pipeline developers in assembling pre-built DS functions and library developers in making these functions available.}
%     \label{fig:users}
% \end{figure}

\paragraph*{RQ 1} Given that the main DS libraries are implemented in Python, their use by programming novices for building pipelines is out of question. Even for experienced programmers, Python libraries that lack type annotations for linters are easy to misuse \cite{reimannImprovingLearnabilityMachine2022}. To make the wealth of DS libraries (Sec. \ref{sec:state-of-the-art}) usable by novices and prevent misuse by users of all skill levels we propose to reuse these libraries via Safe-DS, a DSL for DS that does the heavy lifting. 
% makes it easier to \emph{apply} them in a DS pipeline (RQ 1). 

% TODO: Nach Implementierung verschieben:
% To execute DSL code, we simply transpile it to Python and run the created code in a Python interpreter that contains the Python DS libraries. We call this DSL \keyword{\safeds{}}. 

\begin{figure}[ht]
    \centering
    \includegraphics[width=.48\textwidth]{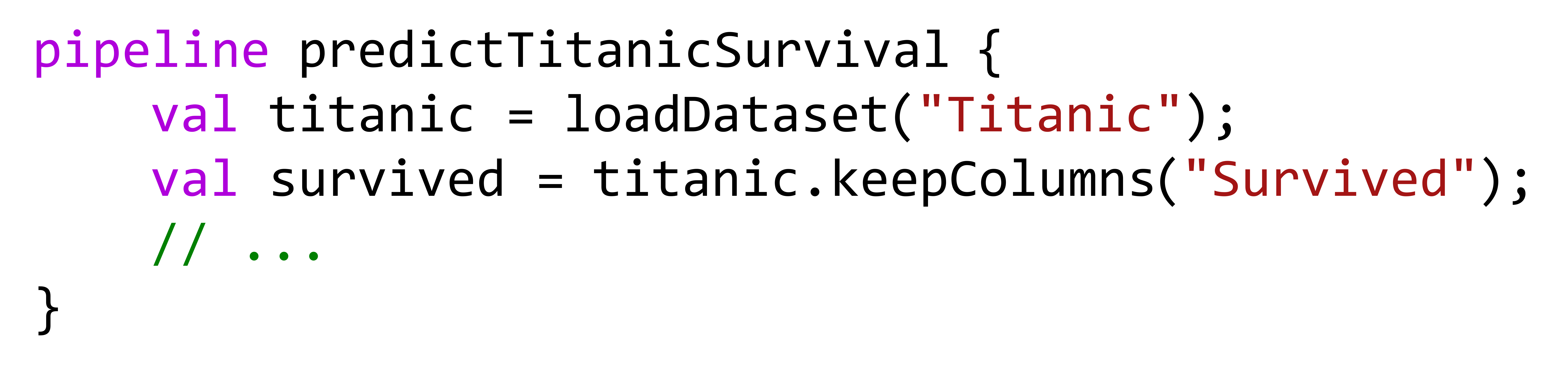}
    \caption{The start of a \safeds{} pipeline to predict survival during the Titanic accident.}
    \label{fig:pipeline-example}
\end{figure}

\paragraph*{RQ 2} This raises the question, which language features are necessary for unrestricted pipeline development and which must be avoided to ease learning and use by novices. Because pipeline development is not about algorithm implementation but mainly a sequence of invocations of already implemented library functions, the set of necessary language features is simple. 
Fig. \ref{fig:pipeline-example} shows the start of a \safeds{} pipeline using the popular ``Titanic'' example. % \cite{Titanic example}: 

The keyword \keyword{pipeline} is the entry point into a \safeds{} program. 
The one in our example is called \code{predictTitanicSurvival}. 
A pipeline contains a sequence of statements, usually assignments\footnote{The only other statement, besides assignments, is the expression statement, which evaluates an expression for its side effects, ignoring  results.}. 
The right-hand side of each assignment contains the expression to evaluate. An expression can be a literal (\code{\str{Titanic}}), a reference to a variable (\code{titanic}), or a call. A call invokes a global function, like \code{loadDataset}, or a method, like \code{keepColumns}. We refer to methods and global functions as \keyword{processes}. Calls can contain further expressions as arguments.

Processes can take lambdas (anonymous, parameterized sequences of statements) as arguments, to configure parts of their behavior. For example, a general \code{transform} function that changes values in one column of a tabular dataset needs to know how to actually change these values. Such higher-order processes are mainly used for data processing functions and rarely needed otherwise.

Notably, in Safe-DS there are no conditionals, loops, or recursive calls. These are encapsulated in the library functions invoked in the pipeline (e.g. hyperparameter search). Decisions and iterations by the developer (reflect on the outcome of the current pipeline, modify it, and restart it) happen outside the pipeline.
% Obiges ist gedacht als Ersatz für folgendes:
% Instead, we offer more concrete, domain specific building blocks, like a hyperparameter search, which hide any conditionals or loops in their Python implementation.
The lack of loops and conditionals makes the language easy to learn.
%enables easier transition between different views (Sec. \ref{sec:different-views}), 
It also eases thorough static checking (Sec. \ref{sec:static-checking}) because data flows linearly through the program.

% ====================================================================
%  Was hier auskommentiert stand ausgelagert nach "backup-xyz.tex"
% ====================================================================

\section{Static Checking}
\label{sec:static-checking}

Supporting safety during the development of a DS pipeline is one of our key aims. For this, detecting errors \emph{statically}, that is before programs are run, is crucial, since the execution of DS pipelines can take hours or even days on large datasets. Achieving static checking, raises the research questions RQ 3 to RQ 6.

\subsection{Type Problems}
\label{sec:type-problems}

% RQ 3: Which classes of errors are not caught by conventional type systems (especially those of Python linters) but occur frequently in DS development

%\paragraph*{RQ 3 -- DS-specific errors} 
The empirical evaluation of \cite{reimannImprovingLearnabilityMachine2022} showed that basic typing features of Python linters already catch many common errors. However, we also identified typical problems that go beyond conventional type systems and are particularly frequent or important in a DS context (RQ 3).

\paragraph*{Lack of range restrictions} We found that the documentation of a library often expresses range restriction, e.g. ratios must be a numeric value between 0 and 1.
% but the most that we can tell a linter is that the value must be a float.  

\paragraph*{Broken dataset access} When processing tabular datasets by adding, removing or renaming columns, it is easy to forget updating some operations that still access a changed column, because column names are just strings and the type system does not know about their special semantics. Neither does it know how to infer the type of columns.  

\paragraph*{Broken order} Machine learning processes must often be called in a specific order (e.g. training before prediction). This simple example could be solved at API level by distinguishing the types \code{UntrainedModel} and \code{TrainedModel}, where only the latter would have a \code{predict} method. However, this would exclude integration of existing libraries that mutate an untrained model in place (e.g. \sklearn{}), would assume redesign of the library, and would bloat the number of types.

%• RQ 4: Which type concepts could help and should be part of a dedicated DS language?

\subsection{Type System}
\label{sec:type-system}

Having identified challenging type problems (RQ 3) we now introduce the solutions provided by Safe-DS (RQ 4).  

% Taking advantage of the intentional simplicity of \safeds{}, 
% particularly the omission of conditionals and loops (Sec. \ref{sec:dsl}), 
% uur solution goes beyond conventional static type systems. In addition to nominal types, genericity, higher-order functions and type inference, it supports inference and checking of dataset schemas and execution of functions in the proper order. 

\paragraph*{Basic type system} 
To capture the basic aspects of DS APIs, the Safe-DS type system supports class types, enum types, function types (which include types for parameters and results), union types, subtyping, and genericity. 

\paragraph*{Refined types} As a way to express range restrictions, Safe-DS
additionally includes refined types, which restrict the legal set of values of an existing type via some constraint. For parameters that have a refined type, only expressions that can be evaluated to a constant value by the compiler can be passed. Otherwise, the compiler would not be able to verify the constraint of the refined type.

\paragraph*{Schema inference and checking} To address the broken dataset access problem (Sec. \ref{sec:type-problems} ) \safeds{} checks correctness of schema-sensitive operations, without having to run the \emph{entire} DS pipeline up to that point:
(1) When a tabular dataset is loaded from external sources, e.g. a CSV file, the compiler infers its initial schema by analyzing the data.
(2) For any data processing function invoked in the pipeline, 
    the compiler inductively infers the schema of the resulting dataset 
    from the schemata of the input datasets and a description 
    of the function's \keyword{schema effects}. 
Our schema effect system \cite{lucassenPolymorphicEffectSystems1988a, marinoGenericTypeandeffectSystem2009a} captures any combination of column addition, removal, renaming, or type change.
In addition, we support two basic schema checks: (1) Ensuring a column with a given name exists. (2) Ensuring a column has a given type. Based on the results of schema inference, the compiler executes such checks statically, thus ensuring early feedback to the user.

\paragraph*{Order checking} 
% Finally, processes must often be called in the correct order. For example, it makes no sense to make predictions with an ML model that has not been trained yet. This can be solved on an API level by differentiating between an \code{UntrainedModel} and a \code{TrainedModel}, where only the latter has a \code{predict} method. However, to ease integration of existing libraries, which often mutate an untrained model in place, we also want to support behavior protocols \cite{plasilBehaviorProtocolsSoftware2002} in the stub language. These are effectively regular expressions that describe legal call orders, with processes instead of characters as basic tokens. Using this information we can then ensure that the call order in pipelines is correct.

%It is often critical that processes are called in the correct order, e.g. we do not want to use an untrained model for predictions.    
%This simple example could be achieved at API level by distinguishing the types \code{UntrainedModel} and \code{TrainedModel}, where only the latter would have a \code{predict} method. However, this would exclude integration of existing libraries that mutate an untrained model in place and would bloat the number of types. 
%
To solve order issues, we support behavior protocols \cite{plasilBehaviorProtocolsSoftware2002} in interface descriptions.
% in the stub language. 
These are regular expressions that describe legal call orders, with processes instead of characters as basic tokens. For instance, we can specify that on any model, the process that trains the model must be called before the one that makes predictions. Using this information, we can ensure statically that the call order in pipelines is correct. 
% , avoiding the limitations of an API-level solution.

\subsection{Integrating External Code}
\label{sec:checking-external-code}
\label{sec:stubs}

%• RQ 5: How can a DS language ensure statically safe use of DS libraries written in dynamic languages?
So far, we have not explained, how the Safe-DS compiler can know at all which functions are available, and how it can statically check calls to functions implemented in a language that does not provide a static type system (RQ 5).

Hard-coding a fixed set of processes into the grammar of \safeds{} would clearly preclude extensibility. This could be avoided by analyzing the code of external libraries, collecting contained type hints or inferring the types of arguments and results while the user develops a pipeline. However, many Python DS libraries do not provide type hints and type inference is not generally possible for Python.

\begin{figure}[ht]
    \centering
    \includegraphics[width=.48\textwidth]{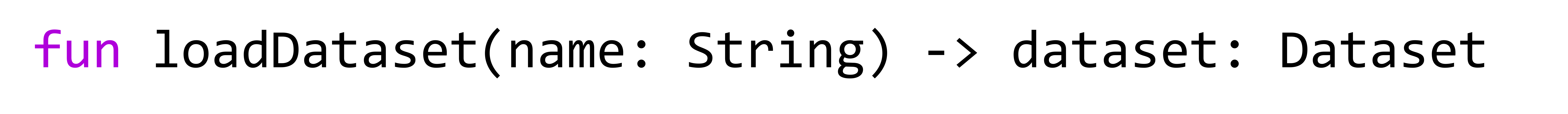}
    \caption{The \safeds{} stub for the \code{loadDataset} function.}
    \label{fig:stub-example}
\end{figure}

Therefore, we designed \safeds{} as a combination of the \emph{pipeline language} described in Sec. \ref{sec:dsl} with an additional \keyword{stub language}. Each stub is an accurate description of the fully typed interface of a particular Python function. 
Fig. \ref{fig:stub-example} shows the stub for the \code{loadDataset} process. The \keyword{fun} keyword indicates that \code{loadDataset} is a function. 
For each parameter (in parentheses) or result (after arrow) we list its name and type, separated by a colon. 

Pipelines can now invoke the specified \code{loadDataset} process (Fig. \ref{fig:pipeline-example}). The \safeds{} compiler checks type-correctness of invocations, by comparing the types of arguments (inferred) with those specified in the respective stub. 
%The compiler uses the closed-world assumption, so 
Any process call for which there is no matching stub is assumed to be wrong, leading to an error message. 

\begin{figure}[ht]
    \centering
    \includegraphics[width=.48\textwidth]{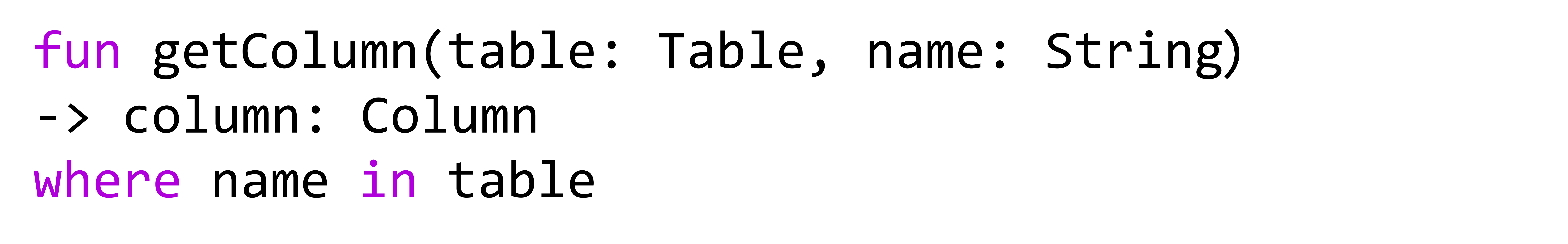}
    \caption{A function stub with a schema check to ensure a column called \code{name} exists in the \code{table}.}
    \label{fig:schema-check-example}
\end{figure}

Stubs tell the \safeds{} compiler all it needs to know for type checking. Therefore, the stub language also includes ways to specify refined types, schema effects, schema checks, and behaviour protocols. For example, the process \code{get\_column} in Fig. \ref{fig:schema-check-example}, which returns the column of the \code{table} with the given \code{name}, specifies that a column called \code{name} must exist in \code{table}. 

\subsection{Semi-Automated Stub Generation and Implementation}
\label{sec:automated-stub-generation}

% RQ 6: How can we make integration of existing Python DS libraries easy, even for third parties?

With the stub language, implementers of new machine learning models or other useful functions can integrate them into \safeds{}. However, integrating an entire library manually would be tedious and a major deterrent to the adoption of \safeds{}. Thus, we need to explain how this process can be eased (RQ 6). 

Our approach adopts and evolves the API-Editor approach of \cite{reimannImprovingLearnabilityMachine2022}.
In a first step, types are inferred for Python code based on type hints in the code \cite{TypingSupportType} or, if type hints are not available, documentation.
For the cases that cannot be inferred automatically,
%, and for review of inferred types,
an easy to use GUI is provided. It also supports specification of changes to the API (e.g. removal of declarations, renaming, and creation of parameter objects). Some of these suggestions can even be inferred automatically. Once the improved and simplified API is finalized, \safeds{} stubs and their corresponding implementations as Python wrappers around the original library are created automatically.

% \discussionafter{Hier wiedersprechen wir uns selbst, denn wir sagten, man könnte keine Typen für Python inferieren. Was wir tun ist aber genau das, nur dass wir sie nicht nur aus dem Code sondern aus der Dokumentation inferieren. Damit hätten wir ein ganz andere, knackigere Story: Typinferenz aus Python Code ist allgemein nicht möglich. Wir nutzen aber die Tatsache, dass keine weitverbreitete Bibliothek ohne eine adäquate Dokumentation daherkommt und extrahieren die Typen aus der Dokumentation (so wie es Reimann et al vorgemacht haben). Damit könne wir dann Stubs generieren, die die Aussenwelt beschreiben. }
%\begin{enumerate}
%    \item Automatically gather information about the original API and its usages.
%    \item \emph{(Optional)} Automatically suggest improvements.
%   \item \emph{(Optional)} Manually review suggested improvements and further alter API.
%    \item Automatically create Python wrappers to implement the improved API.
%\end{enumerate}

This way, large libraries can quickly be integrated into \safeds{}, with only limited manual intervention for cases where the documentation is not accurate enough. 

% \discussionafter{Add results of evaluation to stress that we still need user input and we somehow want to force authors of a library to specify type information, where inference failed to ensure correctness of client programs.}

\section{Different Views}
\label{sec:different-views}

So far, we discussed representing \safeds{} pipelines as text (Sec. \ref{sec:dsl}). However, this is only suitable for people with some programming skills, which does not apply to everyone involved in the development of DS pipelines. 
% For these people the free form of code leads to a large amount of syntax errors. 
To encourage people without programming skills to try solving DS problems, and understand solutions of experts, we provide different views onto DS pipelines, tailored for specific skillsets (RQ 7). 

% Users with a basic grasp of DS might appreciate the flexibility of a drag-and-drop interface to create a graph in a visual DSL consisting of nodes for processes and edges for data flow between them.

\begin{figure}[ht]
    \centering
    \includegraphics[width=.43\textwidth]{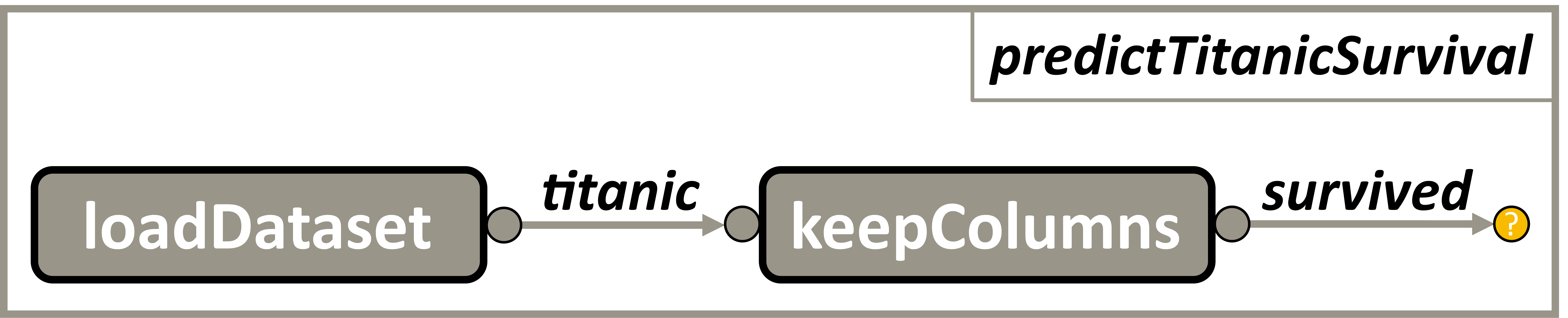}
    \caption{Graphical representation of the pipeline from Fig. \ref{fig:pipeline-example}. Literals are only revealed when clicking on a node, to keep the graph small.}
    \label{fig:graphical_pipeline}
\end{figure}

% Use active voice whenever possible!
%- A possible translation of the pipeline from Fig. \ref{fig:pipeline-example} into a visual DSL is shown in Fig. \ref{fig:graphical_pipeline}. 
Fig. \ref{fig:graphical_pipeline} shows a possible translation of the pipeline from Fig. \ref{fig:pipeline-example} into a visual DSL, which still offers full flexibility but improves safety by preventing syntax errors.
The surrounding box corresponds to the pipeline, with the name shown in the top right corner. The rounded rectangles inside correspond to the invoked processes, with circles for parameters (left) and results (right). This information is retrieved from the stub of the process (Sec. \ref{sec:checking-external-code}). 

Arrows represent dataflow between processes, 
%+
e.g. the arrow from the \code{loadDataset} to the \code{keepColumns} process says that the result of the first process is input to the second one. 
The name shown above an arrow corresponds to the name of a variable in the textual DSL.
The rightmost arrow is attached to a ``blank'' circle since this value is never used but we still want to show the name of the variable that stores it, 
for easier mapping to the textual representation and as a hint to possibly unfinished pipelines.
Literal inputs (e.g. strings) are hidden in the graph representation 
%+ 
to avoid clutter.
They can be made visible by clicking on a process. This approach helps keep the graph small. It is also used in RapidMiner Studio \cite{mierswaYALERapidPrototyping2006}.

%Code and graph views of a pipeline are implemented individually in the state of the art (Sec. \ref{sec:state-of-the-art}). 
Existing systems (Sec. \ref{sec:state-of-the-art}) provide either a code or a graph view of a pipeline, or in case of \cite{keryFutureNotebookProgramming2020} both but only with a fully defined transition from graph to code view. 
We 
%additionally want to be able to 
think that supporting both views, including seamless transitions in both direction, is crucial, to let each person involved in the development of a DS pipeline use the view best suited for them, and even switch between the views, as needed. It would, for example, facilitate collaboration between domain experts and experienced Data Scientists \cite{maoHowDataScientists2019, zhangHowDataScience2020}: A domain expert could prototype an initial pipeline using the visual DSL and then pass it to a Data Scientist, who elaborates and refines it in the textual DSL. Afterwards, the Data Scientist could explain his solution to the domain expert using again the visual DSL, where they could make further changes together.

The switch between different views is only possible if the target view is at least as expressive as the source view. 
%We aim to be able to go from the wizard to the textual or visual DSL and to be able to switch between textual and visual DSL. 
%
Since we want to be able to seamlessly switch between textual and graphical view back and forth, this implies that both representations must be equivalent. Our approach trivially guarantees this, otherwise hard to achieve, prerequisite. In \safeds{}, text and graph are just two different notations for the \emph{same} language.

\section{Prototype}
\label{sec:prototype}

\paragraph*{Language}
We implemented a prototype of \safeds{}\footnote{\url{https://github.com/Safe-DS/DSL}} with Xtext\footnote{\url{https://www.eclipse.org/Xtext}}. At the time of writing, it contains the complete textual syntax for the pipeline and stub languages, scoping, automated formatting of \safeds{} code, generation of Python code, and a vast amount of static checks. 
%This allows it to be used for any view, provided we know which element of the metamodel belongs to which element in the view. 
Static checking currently supports the basic type system without genericity and refined types. It also supports type inference in pipelines. Checking of schema effects is work in progress and behavior protocols are future work. We use Xtext to create a language server \cite{lspcontributorsOfficialPageLanguage}, which we make available for Visual Studio Code\footnote{\url{https://code.visualstudio.com}} via a small extension\footnote{\url{https://marketplace.visualstudio.com/items?itemName=safe-ds.safe-ds}}. 
% The language server could be integrated into any other IDE that support the language server protocol \cite{lspcontributorsOfficialPageLanguage}.

\paragraph*{Metamodel}
The core of the language is the \keyword{metamodel}, the abstract representation of the language elements, also known as the \keyword{abstract syntax} of the language. Individual DS pipelines are instances of this metamodel. The static checks are implemented on top of the metamodel. Each view of the DSL (Sec. \ref{sec:different-views}) defines a transition to and from the metamodel. First, this allows the static analysis to be defined independently from a concrete view, which only needs to display errors to the user in an appropriate form. Second, we avoid having to define and implement transitions between all current and future views. The metamodel is implemented with EMF\footnote{\url{https://www.eclipse.org/modeling/emf}}.

% Complete novices might need the rigid structure of a wizard, potentially even using AutoML methods to make most of the steps of the wizard optional. 
% \discussionafter{Ich würde die wizards und Auto-ML weiter nach hinten schieben, da sie hier den Argumentations fluss von Text zu graphen unterbrechen und die Wizards sowieso nicht etwas sind, was durch unsere DSL ermöglicht wird. WIr sollten hier auf die Essenz Fokussieren (de Äquivalenz der textuellen und graphischen Srache). } 

\begin{figure}[ht]
    \centering
    \includegraphics[width=.40\textwidth]{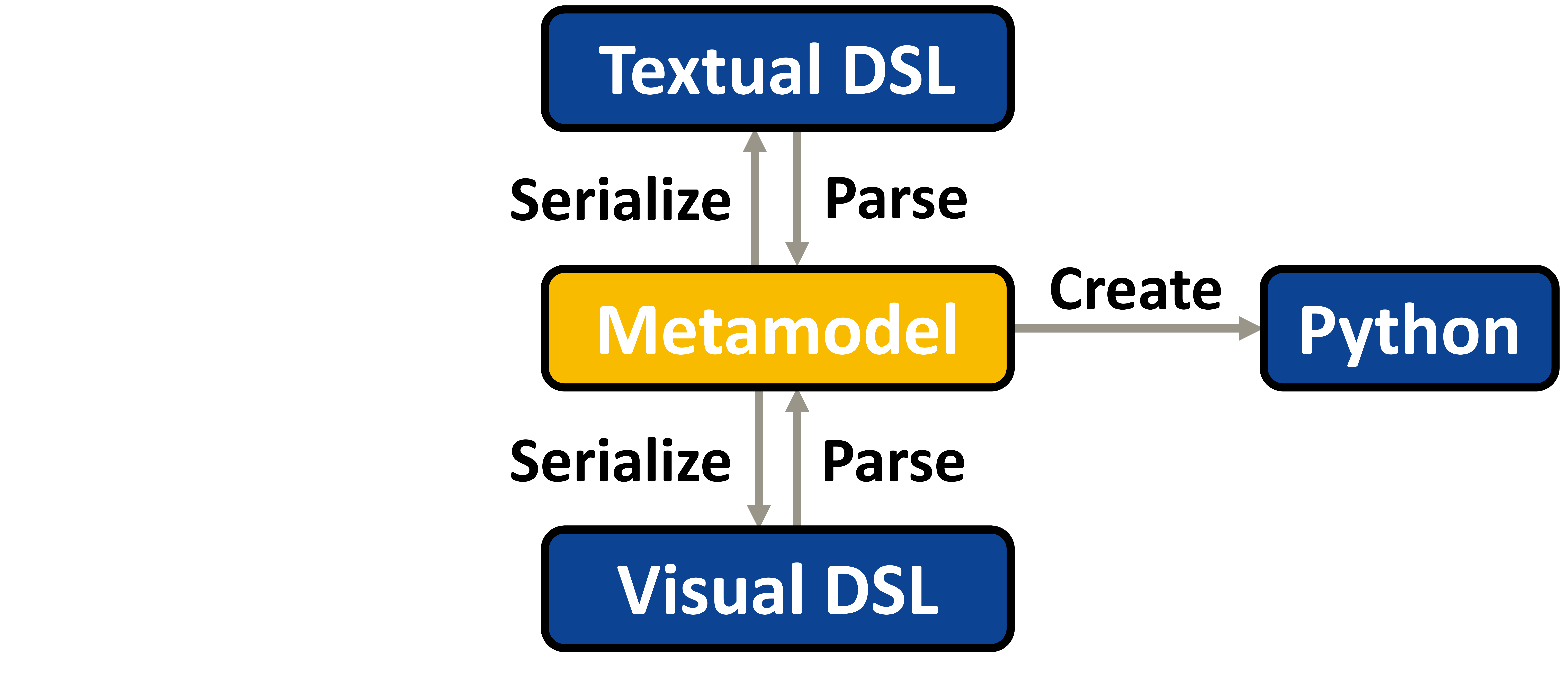}
    \caption{Interaction between different views (blue) with the metamodel to connect them.}
    \label{fig:different-views}
\end{figure}

Fig. \ref{fig:different-views} shows the relation between the textual and graphical view and the metamodel. It also shows the one-way transition from the metamodel to Python code generated by the compiler.

% First, we can support wizards that guide complete novices through the creation of a pipeline and generate a \safeds{} program. This can be executed directly, or can be further improved or extended in either representation, textual or visual.

% \discussionbefore{AutoML würde ich weglassen, da es unabhängig von unserem Ansatz immer möglich ist, also kein Argument ist, das uns besonders weiter hilft. Wir müssen nicht die eierlegende Wollmilchsau vorschlagen nur um unseren Ansatz interessant zu machen. Schon der Wizard ist nicht essentiell, aber er passt zumindest, genauso die Idee nach Python zu exportieren. }
% Potentially even using AutoML methods to make most of the steps of the wizard optional. 
%
% To support the other end of the skills spectrum, the translation of \safeds{} to Python, which exists anyway to execute a \safeds{} pipeline, can be made available to developers as a basis for inclusion in pure Python environments. It only poses the additional requirement that the generated Python code should be readable.

\paragraph*{Standard library}
Moreover, we started work on the Safe-DS standard library. So far, it includes basic functionality to work with tabular data (using pandas \cite{The_pandas_development_team_pandas-dev_pandas_Pandas} internally), visualize this data (using seaborn \cite{Waskom2021}), and train and evaluate ML models (using scikit-learn \cite{scikit-learn, sklearn_api}). An experimental version of the corresponding Python implementation is available on PyPI\footnote{\url{https://pypi.org/project/safe-ds}}. We will discuss the standard library in more detail in a future publication.

%GK ⚠️ Discuss how feasible it is to annotate the various Python libraries that should be supported
%  * We are doing this already for the core parts of DS libraries
%  * We will report how much effort it was once that process is done
%  * Others can use the same mechanism to integrate additional features. We think the effort is relatively low.

\section{Discussion}
\label{sec:discussion}

\paragraph*{Accuracy of stubs}
The static analysis requires that stubs accurately describe the preconditions of the corresponding Python implementation. A mismatch can either cause valid DSL programs to be rejected or invalid DSL programs to be executed, leading to runtime exceptions or undefined behavior. We aim to reduce the chance of such mismatches by automatically creating stubs for Python declarations where possible (Sec. \ref{sec:automated-stub-generation}). Furthermore, users only need to work with the stub language when they want to integrate functionality that is missing from the standard library. For the standard library itself, we pay particular attention to the accuracy of the stubs.

\paragraph*{Omission of language concepts}
While the language itself contains neither loops nor conditionals, nothing stops a user from implementing higher-order \code{if} or \code{while} functions in Python and integrating these into \safeds{} with its stub language. We would treat these like any other function, only using the information from the stubs for static analysis or the representation in the graphical view. On the one hand, this means we ignore the semantics of the if and while constructs for static analysis. On the other hand, such constructs are not needed in the first place since we offer domain specific abstractions like a \code{transform} function for tabular datasets, which encapsulate any loops or conditionals in the Python implementation and use features of the DSL like the schema checking (Sec. \ref{sec:type-system}) for domain specific validation.

% \todo{
% \begin{itemize}
%     \item A discussion about the safety of the stubs is missing. The reviewer can easily imagine a situation where a stub wrongly reflects the Python function and is called deep in the code such that the crash can happen days into running the pipeline.
% \end{itemize}
% }

\section{Future Plans}
\label{sec:future-plans}

\paragraph*{Additional checks} We are working on completing the implementation of schema checking and will then add checking of call order using behavior protocols. We also look into whether we can extend \cite{reimannImprovingLearnabilityMachine2022} to partially infer schema effects and behavior protocols automatically from documentation.

\paragraph*{Standard library} In parallel, we are integrating additional features from Pandas \cite{The_pandas_development_team_pandas-dev_pandas_Pandas} for data preparation, seaborn for visualization \cite{Waskom2021}, and \sklearn{} \cite{scikit-learn, sklearn_api} for traditional machine learning, using the already implemented stub language (Sec. \ref{sec:checking-external-code}) and automated stub generation (Sec. \ref{sec:automated-stub-generation}). 
%Later, libraries for Deep Learning, e.g. PyTorch \cite{NEURIPS2019_9015}, can be added, too.

\paragraph*{Usability study} When a sufficient part of the standard library is implemented, we will conduct a quantitative and qualitative comparison of \safeds{} and Python for solving real world problems. 

\paragraph*{Different views} Currently, the available Visual Studio Code extension for \safeds{} only has a textual view. We plan to also implement a graphical view (Sec. \ref{sec:different-views}) using GLSP\footnote{\url{https://www.eclipse.org/glsp/}}. 
%
% Later, we also plan to add a wizard to help users create an initial DS pipeline, which can then be altered in the textual or graphical view.

% \todo{
%     \begin{itemize}
%         \item Improve stub generation (e.g. behavior protocols)
%         \item ---
%         \item Further enhance Xtext prototype
%         \item Further enhance VS code extension
%     \end{itemize}
% }

\section{Conclusions}
\label{sec:conclusion}

In this paper we focused on the needs of DS applications, rather than DS libraries. We presented \safeds{} (Sec. \ref{sec:dsl}), a %textual 
DSL for the development of DS pipelines. It combines the expressive power of Python DS libraries with safety via compile-time error detection.

Type-safe integration of Python code is based on creation of stubs (Sec. \ref{sec:checking-external-code}) that describe 
the interface of external functions, following largely automated type extraction from a library's code \emph{and} documentation (Sec. \ref{sec:automated-stub-generation}). 
The current implementation of \safeds{} already supports a full conventional static type system, with manifest types in the stubs, and type inference in DS pipelines. 
%In addition, we have pointed out two classes of errors that cannot be caught by conventional type systems
%and are particularly nasty in a DS contexts, because an extreme amount of time and resources is often wasted before they manifest themselves by run-time crashes or nonsense results. 
%They concern the necessary data preprocessing and machine learning. 
Adaptations of non-standard type concepts, such as effect systems (to check schema-conformance of operations on tabular datasets) and behaviour protocols (to check call order) can help deal with these issues statically, too (Sec. \ref{sec:type-system}). 
% We use an effect system for checking schema-sensitive operations during processing of tabular, and behavior protocols to validate the call order of processes in a pipeline.

% The DSL code is transpiled to Python for execution. This also allows users to opt out of the system, 
% should the DSL prove too restrictive for a use case. \discussionafter{Der letzte Nebensatz gefällt mir nicht, denn er suggeriert, dass es Dinge gibt, die wir nicht ausdrücken können, was ja der Behauptung, dass man Schliefen und Conditionals nicht braucht widerspricht. Entweder braucht man sie oder man braucht sie nicht... Ich könnte mir eher eine Argumentation vorstellen, dass man aus welchen Gründen auch immer unbedingt eine reine Python-Lösung haben will / muss, oder manche Benutzer meinen, das sie keine Hilfe brauchen um korrekte Programme zu schreiben. }

%In order to enable people with limited programming skills to develop DS pipelines, 
To enable DS pipeline development by people with limited programmings skills 
and foster collaboration of people with different skill levels, 
%we eventually want to support other views onto the same DS pipeline (Sec. \ref{sec:different-views}). 
our approach supports seamless transition between textual and graphical incarnations of the language (Sec. \ref{sec:different-views}). 
% These views can include a wizard that guides novices through the development process 
% or a visual DSL that allows the development of more flexible DS pipelines without having to write code. 
% We also want to have well-defined transitions between different views via the metamodel of the language 
% so people with different skills can work on the same DS pipeline using the view most suited for them.
%
%More flexible views, however, also have higher potential for errors. 
%We aim to catch these errors statically with minimal interaction with 
%the developer of a DS pipeline (Sec \ref{sec:static-checking}). 
%Our static checking will include a type system to detect basic type errors, 
%a schema checker to ensure correct processing of tabular data, and behavior protocols to validate the call order of processes in a pipeline.
%
Overall, we hope that \safeds{} will prove valuable to help novices
%, students and domain experts,
get started with Data Science.

\section*{Acknowledgments}

This work was partially funded by the Federal Ministry of Education and Research (BMBF), Germany under the Simple-ML project (grant 01IS18054). We are grateful for all former contributions to the Safe-DS GitHub repository, particularly with respect to schema checking and the standard library. We also thank the reviewers for their numerous constructive comments and suggestions.

% Bibliography ---------------------------------------------------------------------------------------------------------

\newpage

\bibliographystyle{IEEEtran}
\bibliography{IEEEabrv,references}

\end{document}